\documentclass[
aps,%
10pt,%
final,%
notitlepage,%
oneside,%
twocolumn,%
nobibnotes,%
nofootinbib,%
superscriptaddress,%
noshowpacs,%
centertags]%
{revtex4-2}

\usepackage{amsmath,amssymb,bm}
\usepackage{array,tabularx}
\allowdisplaybreaks[3]
\usepackage{graphicx}\graphicspath{{}{Graph1/}{Graph/}}
    \usepackage[svgnames]{xcolor}

    \usepackage[T1]{fontenc} 
    \usepackage[utf8]{inputenc} 

\usepackage[russian, english, showlanguages]{babel}

    \DeclareMathOperator{\e}{e}

    \newcommand{\parder}[2]{\frac{\partial #1}{\partial #2}}

    \newcommand{\dif}[2][]{\mathop{}\!\mathrm{d}
        \if
            \relax\detokenize{#1}\relax
        \else
            ^{\mkern-1.mu#1}\mkern-2.5mu 
    \fi
    #2\,}

\usepackage[
    colorlinks
        ,linkcolor=violet
        ,filecolor=purple
        ,citecolor=teal
        ,urlcolor=magenta
    ,unicode
    ,pdfpagemode=UseOutlines
    ,pdfdisplaydoctitle=true
    ,pdfpagetransition=Wipe
    ,pdfusetitle
]{hyperref}
\hypersetup{
        pdfkeywords={plasma, plasma diagnostics}
}
\usepackage{bookmark}

\begin{document}

\title{Theory of a dynamic plasma flow pressure sensor}

\author{\firstname{Evgeny} \surname{Kolesnikov}}
\email{E.Yu.Kolesnikov@inp.nsk.su}
\author{\firstname{Igor} \surname{Kotelnikov}}
\email{I.A.Kotelnikov@inp.nsk.su}
\author{\firstname{Vadim} \surname{Prikhodko}}
\email{V.V.Prikhodko@inp.nsk.su}
\affiliation{Budker Institute of Nuclear Physics, Novosibirsk, Russia}
\affiliation{Novosibirsk State University, Novosibirsk, Russia}
\date{\today}

\begin{abstract}
The problem of reconstructing the time dependence of the dynamic pressure of a plasma jet impinging on one end of a solid rod based on the measured displacement of the opposite end has been solved. This solution allows for a reduction in the size of the dynamic pressure sensor proposed and later improved in the works \cite{Kostukevich2002PublAstronObs_74_149, Astashynski+2014PPT_1_157}.
\end{abstract}

\maketitle

\section{Introduction}
\label{s1}


At the Budker Institute of Nuclear Physics (INP SB RAS), experiments are being conducted to develop methods for maintaining the material balance of plasma in an open-type Gas-Dynamic Trap (GDT) \cite{IvanovPrikhodko2017PhysUsp_60_509}. One of these methods is to measure the pressure of the plasma flow injected into the GDT from a coaxial plasma accelerator (CPA), also known as the Marshall gun \cite{Morozov2008}. The CPA consists of two coaxial cylindrical electrodes, between which a pulsed current of the order of $J\approx100\,\text{kA}$ flows in hydrogen plasma with a characteristic rise time of about $\tau\approx 10\,\mu\text{s}$. Closed through the central electrode, the current creates an azimuthal magnetic field. When the radial current through the plasma interacts with the azimuthal magnetic field, the plasma is accelerated under the influence of the Ampere force. The characteristic parameters of the accelerated plasma are as follows: density $n\sim10^{15}\text{–}10^{16}\,\text{cm}^{-3}$, velocity $u\sim10^7\,\text{cm/s}$, and the pressure $p$ created by the directed plasma progeny can reach several atmospheres.


Measuring the parameters of accelerated plasma is complicated by the high instantaneous power density of the resulting jet ($P>0.5\,\text{GW}/\text{m}^2$) and significant interference generated by the high pulsed current. A promising method for measuring the pressure of a powerful plasma jet is the dynamic pressure detection method \cite{Kostukevich2002PublAstronObs_74_149, Astashynski+2014PPT_1_157} developed by a group of authors from the Academy of Sciences of the Republic of Belarus.

\begin{figure}
  \centering
  \includegraphics[width=\linewidth]{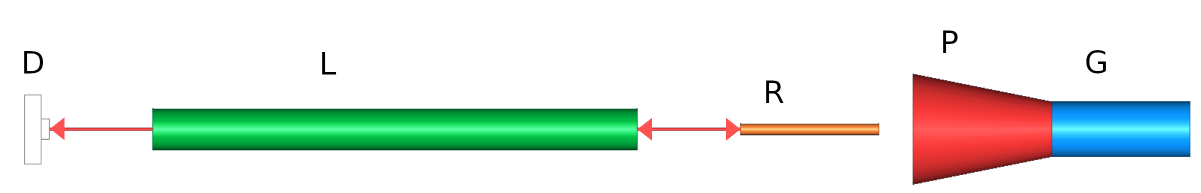}
  \caption{
Schematic diagram of the plasma flow dynamic pressure sensor:
D—laser intensity detector,
L—helium-neon laser,
R—measuring rod,
P—plasma jet,
G—plasma source.
  }\label{fig:Kolesnikov}
\end{figure}

A plasma jet from the CPA is directed at the end of a copper cylindrical rod, the diameter of which is much smaller than the length $l$, as shown in Fig.~\ref{fig:Kolesnikov}. A mirror is deposited on the flat end of the rod opposite the plasma. The mirror reflects the laser beam directed on it. The impact of the plasma stream on the right end excites a compression wave, which propagates along the rod at the speed of sound $c$, causing a displacement of the left end of the rod opposite the plasma after a time of $l/c$, which is required for the sound wave to travel from one end of the rod to the other. This displacement creates a phase shift in the reflected laser beam and is recorded using an optical detector. In the paper \cite{Astashynski+2014PPT_1_157}, it is proposed to use a He-Ne laser active medium coaxial with the rod for this purpose. The laser power modulation will correspond to the phase shift due to the movement of the mirror end of the rod.

A shift of the rod end by half the laser wavelength $\lambda $ corresponds to a full period of laser power modulation. By measuring the time between maxima of the optical signal intensity, the velocity of the rod end $v(t)$ can be determined. Further, based on the end velocity $v(t)$, the speed of sound $c$, and the density of the rod material $\rho$, one can calculate the plasma pressure force $f(t)$ on the rod end facing the CPA, and even determine how this pressure changes over time.


In this paper, we present the theory of the measurement process in the simplest case, where the influence of the measuring rod fastening elements is insignificant. We demonstrate that the authors of the method \cite{Kostukevich2002PublAstronObs_74_149, Astashynski+2014PPT_1_157} made a mistake by using a formula that relates the velocity and pressure in a traveling sound wave in an infinite medium to calculate the pressure from the displacement velocity of the measuring rod end. We present the correct solution to the inverse problem, expressing the plasma pressure (the cause of the wave) in terms of the end velocity (the result of the wave action). We demonstrate that the resulting solution eliminates the need to use long rods, contrary to the recommendations of the method's authors.


In the next section \ref{s2}, the method for measuring the plasma flow pressure is described in the authors' interpretation, with some clarifications from us. Further, in section \ref{s3}, the direct problem is formulated and solved. In section \ref{s4}, the frequencies of the rod's natural longitudinal oscillations are calculated. In section \ref{s5}, the rod's oscillations under the action of an external force acting on one of the ends are considered. In section \ref{s6}, the inverse problem is solved, and it is shown how to reconstruct the time dependence of the external force from the results of measuring the displacement velocity of the opposite end. In section \ref{s9}, our conclusions are formulated. Appendix \ref{A1} presents an alternative solution to the direct problem using the ``school'' method of tracking multiple reflections of a sound wave from the ends of the rod. A similar method was used in the work \cite{ManzhosovMartynova2021VestnikUlTU_3_86}, but for the special case when the force acting on the end of the rod remains constant after the start of its action.

\section{Dynamic pressure sensor}
\label{s2}


In the Belarusian physicists' method, the dynamic pressure of plasma flows is measured using an interferometric pressure sensor, as shown in Fig. ~\ref{fig:Kolesnikov}. The sensor comprises an acoustic element consisting of a long copper rod, a helium-neon laser, and a photomultiplier. The carefully polished left (rear) end surface of the rod reflects the light beam back into the laser cavity. The intensity of the beam emerging from the left (rear) mirror of the laser and incident on the photomultiplier depends on the phase relationship between the reflected beam and the beam in the cavity.


Upon impact with the right (front) end surface of the rod, the plasma flow generates a compression wave that propagates along the rod at the speed of sound in copper. Reflection of the wave from the right end causes it to shift from its initial position. As a result, the phase relationship between the incident and reflected laser beams, and the intensity of the beam incident on the photomultiplier, is modulated with a frequency proportional to the velocity of the right end displacement. To calculate the plasma flow pressure on the right end, the authors of the method use the formula
    \begin{gather}
    \label{01:01}
    p(t) = c\,\rho\,v(t)
    \qquad 
    \text{[wrongly]}
    ,
    \end{gather}
%
which, in their opinion, establishes the relationship between the instantaneous values of the flow pressure $p(t)$ and the velocity of displacement of the left end $v(t)$. Here $c$ and $\rho$ are the speed of sound in the rod and the density of the rod material, respectively.

%
One period of signal modulation corresponds to a shift equal to half the wavelength of laser radiation, therefore
    \begin{gather}
    \label{01:02}
    v(t) = 0.5\lambda /T
    ,
    \end{gather}
%
where $T$ is the modulation period of the signal arriving at the photomultiplier at time $t$. According to the authors \cite{Astashynski+2014PPT_1_157}, the measurement accuracy is limited by variations in the speed of sound in the rod due to changes in the rod's density during its deformation, which do not exceed 10\%.

%
The key feature of the Belarusian system is the arrangement of all its components on a single optical axis. This ensures ease of setup, reasonable vibration resistance, and signal recording after optical and electrical noise has significantly weakened or ceased (depending on the length of the acoustic element). According to the authors of the method \cite{Astashynski+2014PPT_1_157}, the rod length $l$ should be selected based on the expected duration $\tau$ of the process being studied:
    \begin{gather}
    \label{01:03}
    l \geq c\,\tau/2
    .
    \end{gather}
%
They feared that by the time $t=\tau$ the compression wave, having reflected from the left end, would return back to the right end, causing its additional displacement and thereby distorting the longer signal.

\subsection{Known results and their interpretation}\label{s2.1}

\begin{figure}
  \centering
  \includegraphics[width=\linewidth]{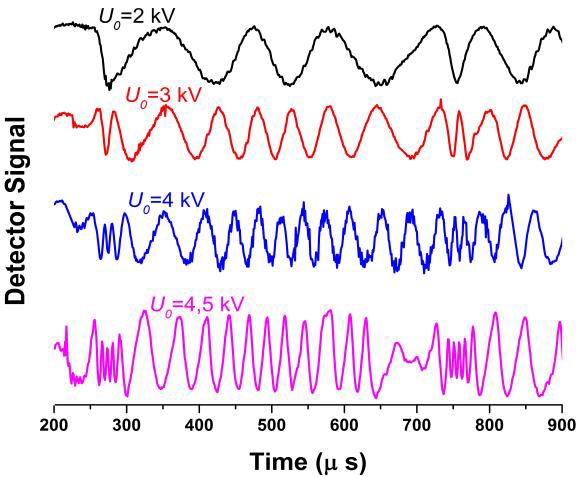}
  \caption{
%
    Copy of Fig. 2 in the article \cite{Astashynski+2014PPT_1_157}:
    Detector signal versus time for voltages $U_{0}=2, 3, 4, 4.5\,\text{kV}$ at the plasma source.
  }\label{fig:Astashynski+2014PPT_1_157-Fig2}
\end{figure}

%
The interference signals published in the article \cite{Astashynski+2014PPT_1_157} are shown in Fig.~\ref{fig:Astashynski+2014PPT_1_157-Fig2}. Apparently, these signals in the theory presented below should be interpreted as
    \begin{gather}
    \label{2.1:01}
    \text{Detector signal}
    \propto
    \sin(4\pi x(t')/\lambda)
    ,
    \end{gather}
%
where $x(t)$ is the displacement of the left end, $\lambda$ is the laser wavelength, and $t'=t-l/c$ is the time measured from the moment the leading edge of the compression wave arrives at the left end of the measuring rod. The fact that the signal has the shape of an almost perfect sinusoid for approximately $500\,\mu\text{s}$ means that the displacement of the end $x(t')$ increases monotonically and almost linearly with time throughout this time.


As soon as the sound wave reaches the left end from the right one after a time of $l/c$ from the start of the plasma flow action on the right end, it immediately generates a reflected wave, therefore the formula \eqref{01:01} is not suitable for calculating the pressure, since it expresses the relationship between the pressure and the displacement velocity in a traveling wave (see~\cite[\S64, Eq.~(64.11)]{LLVI}), whereas at the left end of the rod at any moment $t>l/c$ we have a superposition of the incident and reflected sound waves.


According to the data given in the article \cite{Astashynski+2014PPT_1_157} ($l=0.84\,\text{m}$), and the speed of sound in a copper rod according to the data of \href{https://www.engineeringtoolbox.com/sound-speed-solids-d_713.html}{The Engineering Toolbox} ($c\approx 5\,000\,\text{m}/\text{sec}$), the travel time of a sound wave from end to end of the measuring rod is estimated as $l/c=170\,\mu\text{s}$. It is significantly shorter than the duration of the recorded interference signal $T\approx 900\,\mu\text{s}$ in Fig.~\ref{fig:Astashynski+2014PPT_1_157-Fig2}, but longer than the duration of the pressure pulse $\tau\approx 90\,\mu\text{s}$.


According to the authors of the method, due to this rod length, the sensor could register undistorted signals with a duration of approximately $500\,\text{s}$. This duration is approximately equal to the duration of the regular sinusoid in Fig.~\ref{fig:Astashynski+2014PPT_1_157-Fig2} between the segments of fast sinusoids at times $ t=270\,\mu\text{s}$ and $t=770\,\mu\text{s}$. On the other hand, $500\,\mu\text{s}$ is approximately equal to twice the travel time of the sound wave along the measuring rod. Apparently, the authors of the method call the unperturbed signal the signal that was reflected from the left end, then from the right, but did not have time to reach the left end again, where the measured flow is realized.


The authors of the method estimate the duration of the pressure pulse that generated the signal shown in Fig.~\ref{fig:Astashynski+2014PPT_1_157-Fig2} to be approximately $50\text{-}70\,\mu\text{s}$, that is, approximately $(0.1\text{-}0.2)\,l/c$. We will take this circumstance into account further when constructing an example in section \ref{s5}. The duration of the measured signal $900\,\mu\text{s}$ was more than $5\text{-}6$ times longer than the sound wave travel time $170\,\mu\text{s}$. This is also taken into account when constructing the graphs.

\section{Direct problem}
\label{s3}

Let us have a rod of length $l$, stiffness $k$, and mass $m$. Following Ref.~\cite{Milstein2024Arxiv_2405_00050}, we write the wave equation for the displacement of the rod elements
\begin{gather}
\label{1:0}
\parder{^{2}x}{t^{2}} = c^{2}\parder{^{2}x}{z^{2}}
,
\end{gather}
where
\begin{gather*}
c=\sqrt{{k}/{m}}\,l.
\end{gather*}
On the other hand, in the theory of vibrations of thin rods, the speed of sound is expressed through Young's modulus $E$ and density $\rho$ according to the formula \cite[\S25]{LLVII}
\begin{gather*}
c=\sqrt{{E}/{\rho}}.
\end{gather*}
To this, we must add the equation
\begin{gather*}
m = \pi a^{2}l\rho
,
\end{gather*}
which relates the mass of the rod $m$, its density $\rho$, radius $a$, and length $l$.

To relate the spring stiffness coefficient $k$ to the rod parameters, we take into account that on its end surface the corresponding component of the stress tensor $\sigma_{zz}$ is equal to the pressure $p$:
\begin{gather*}
\sigma_{zz}=p.
\end{gather*}
The $u_{zz}$ component of the displacement tensor, on the one hand, is equal to
\begin{gather*}
u_{zz} = \sigma_{zz}/E
,
\end{gather*}
and on the other hand, it has the meaning of the relative elongation of the rod, that is,
\begin{gather*}
u_{zz} = x(l,t)/l
,
\end{gather*}
where $x(l,t)$ is the displacement of the right end of the rod from the equilibrium position, provided that the left end of the rod is fixed, and therefore $x(0,t)=0$. If the rod is considered as a spring, then
\begin{gather*}
\pi a^{2} p = k\,x(l,t)
.
\end{gather*}
Therefore,
\begin{gather}
\label{01:11}
k = \pi a^{2}\frac{E}{l} = \pi a^{2}\rho\frac{c^{2}}{l}
.
\end{gather}
This formula, of course, is not change even if we use different boundary conditions at the ends of the rod than in its derivation above.

In an infinite medium, the general solution of the wave equation \eqref{1:0} is conveniently sought in the form of two plane waves traveling in opposite directions, with the profile of each wave preserved due to the neglect of sound dispersion in the dispersion equation. In a finite-length rod, this approach requires careful tracking of multiple reflections of these two waves from both ends of the rod, as done in Appendix \ref{A1}. While this approach does not appear to be productive when solving more complex problems, such as calculating the eigenfrequencies of a suspended rod (which will be the subject of a future article), it at least provides confidence in the validity of the theory presented below.

In accordance with the real situation, we assume that the rod is initially at rest, that is, the displacement $x(z,t)$ and velocity $v(z,t)$ in an arbitrary cross-section $z$ of the rod are equal to zero at $t\leq0$:
\begin{gather}
\label{1:1}
x(z,0) = 0,
\\
\label{1:2}
v(z,0) = 0.
\end{gather}
Starting at time $t=0+$, the force $f(t)$ begins to act on it from the right. Formally, this fact describes the boundary condition \cite{Milstein2024Arxiv_2405_00050}
\begin{gather}
\label{1:3}
kl \,\parder{x}{z}(l,t) = f.
\end{gather}
The left edge is free, that is,
\begin{gather}
\label{1:4}
kl\, \parder{x}{z}(0,t) = 0.
\end{gather}
To fully solve the problem, we need to find the displacement of the rod element $x(z,t)$ from the initial position $z$ at each moment of time $t$ and the velocity of this element $v(z,t)=\parder{x}{t}(z,t)$. We only need to calculate the quantities measured in the method under study, that is, the displacement and velocity of the left end of the rod. We will denote these quantities with the same letters, but with one argument: $x(t)=x(0,t)$ and $v(t)=v(0,t)$, or without arguments at all: $x=x(0,t)$ and $v=v(0,t)$.

%
%
%

\section{Eigenfrequencies of Oscillations}
\label{s4}

From the boundary condition \eqref{1:4} it follows that the solution of the wave equation with a certain frequency $\omega_{j}$ is a standing wave:

\begin{gather}
\label{3:1}
x_{j}(z,t) = A_{j}\sin(\omega_{j}t+\alpha_{j})\cos(q_{j}z)
,
\end{gather}
where $q_{j}=\omega_{j}/c$ is the wave number, $A_{j}$ and $\alpha_{j}$ are the amplitude and phase of the wave. Assuming that the right end of the rod is also free, we set the zero boundary condition at $z=l$
\begin{gather}
\label{3:4}
kl\, \parder{x}{z}(l,t) = 0,
\end{gather}
analogous to Eq.~\eqref{1:4}. Substituting \eqref{3:1} into \eqref{3:4} yields the equation \begin{gather}
\label{3:5}
A_{j}\sin(\omega_{j}t+\alpha_{j})\sin(q_{j}l)
=
0.
\end{gather}
This equation has an infinite set of eigenwave numbers
\begin{gather}
\label{3:6}
q_{j} = (\pi/l)\,j
,
\end{gather}
and, consequently, an infinite set of eigenfrequencies
\begin{gather}
\label{3:7}
\omega_{j}=(\pi\,c/l)\,j
,
\end{gather}
where $j=0,1,2,\ldots $. An oscillation with zero frequency $\omega =0$ represents a rigid displacement of the rod without internal deformations.

\section{Forced Oscillations}
\label{s5}

We apply the Laplace transform to the wave equation \eqref{1:0}, that is, we calculate the Mellin integrals for all its terms,
\begin{gather}
\label{05:01}
Y(s,z) =
\int_{0}^{\infty }\e^{-st}y(z,t)\dif{t}
,
\end{gather}
where the function $y(z,t)$ can be replaced by any desired function; from here on, capital letters are used to denote the Laplace image of the original function, which is called the original. Integrating by parts the right-hand side of equation \eqref{1:0} twice under the assumption that $\e^{-st}\to0$ as $t\to+\infty $, we obtain an ordinary differential equation
\begin{gather}
\label{05:02}
-\parder{x}{t}(z,0)
+
s\,x(z,0)
+
s^{2} X(s,z)
=
c^{2}\parder{^{2}X}{z^{2}}(s,z)
.
\end{gather}
Given the zero initial conditions \eqref{1:1} and \eqref{1:2}, its solution is the sum of two exponentials
\begin{gather}
\label{05:03}
X(s,z) = C_{1}\exp(-sz/c) + C_{2}\exp(sz/c)
.
\end{gather}
One of the coefficients $C_{1}$, $C_{2}$ for these exponentials can be expressed in terms of the other coefficient by substituting this solution into the boundary condition \eqref{1:4} on the left boundary $z=0$, so that
\begin{gather}
\label{05:04}
X(s,z) = X(s)\cosh(sz/c)
.
\end{gather}
The solution to the original wave equation yields the inverse Laplace transform, that is, the Bromwich integral (also known as the Fourier-Mellin integral or the inverse Mellin formula)
\begin{gather}
\label{05:05}
x(z,t)
=
\frac{1}{2\pi i}\int\limits_{-i\infty +\gamma }^{i\infty +\gamma }
\e^{st}
X(s)
\cosh(s z/c)
\dif{s}
.
\end{gather}
The Mellin integral \eqref{05:01} is known to converge if the parameter $s$ has a positive and sufficiently large real part. Accordingly, the integration contour in \eqref{05:05} must pass to the right of all singularities of the integrand, meaning the parameter $\gamma $ must not only be positive, $\gamma >0$, but also sufficiently large. This choice of contour ensures that the integral vanishes at $t\to-\infty$, when the external force on the rod is still absent.

Substituting \eqref{05:05} into the boundary condition \eqref{1:3}, we obtain the equation
\begin{gather}
\label{05:06}
f(t)
=
\frac{1}{2\pi i}
\int\limits_{-i\infty +\gamma }^{i\infty +\gamma }
\e^{st}
s\,X(s)
\sinh(sl/c)
\dif{s}
,
\end{gather}
from which we can find the image function $X(s)$. Applying the Mellin integral to both sides of the last equation, we obtain on its left-hand side the image
\begin{gather}
\label{05:07}
F(s)
=
\int\limits_{0}^{\infty }\e^{-st}\,f(t)
\dif{t}
\end{gather}
of the force $f(t)$ acting on the right end of the rod. On the right-hand side, taking into account the identity of the type
\begin{gather}
F(s)
=
\int_{0}^{\infty}
\e^{-st}
\left[
\frac{1}{2\pi i}
\int_{-i\infty+\gamma }^{i\infty+\gamma }
\e^{s't}
F(s')
\dif{s'}
\right]
\dif{t}
\end{gather}
the integrand will be extracted. As a result, we obtain the desired formula:
\begin{gather}
\label{05:09}
F(s)
=
s\,X(s)\sinh(sl/c)
.
\end{gather}
Combining this with the Mellin integral \eqref{05:05}, we find the displacement of the left end of the rod:
\begin{gather}
\label{05:11}
x(t)
=
x(0,t)
=
\int\limits_{-i\infty +\gamma }^{i\infty+\gamma }
\e^{st}\frac{F(s)}{s\sinh(sl/c)}
\dif{s}
.
\end{gather}
The integrand contains an infinite number of poles located on the imaginary axis. They correspond to the excitation of eigen oscillations. The integration contour must go around all the poles on the right. If the function $F(s)$ is entire (that is, regular in the entire complex plane), there are no other poles.

A second-order pole $s=0$ generates a term that grows linearly with time. In the physics of the case, it describes uniform translational motion with constant velocity. This term should disappear if we take into account that the rod is attached to a pair of suspensions.

To simplify the intermediate formulas in the rest of the article, we will further assign the half-period of the sound wave travel $l/c$ between the ends of the rod to the unit of time, retaining the same notations for the dimensionless quantities $t$ and $\tau$. This non-dimensionalization is achieved by substituting $l/c\to1$. If desired, the dimension can be restored later by substituting $t\to ct/l$ and $\tau\to c\tau/l$. Similarly, we will define the unit of force $f$ as the parameter $kl=mc^{2}/l$. To do this, simply omit the factor $kl/c$ or $c/(kl)$ in Eqs.~\eqref{05:06}, \eqref{05:09}, \eqref{05:11}, and similar ones.

\subsection{Example}
\label{s5.1}

\begin{figure}
\centering
\includegraphics[width=\linewidth]{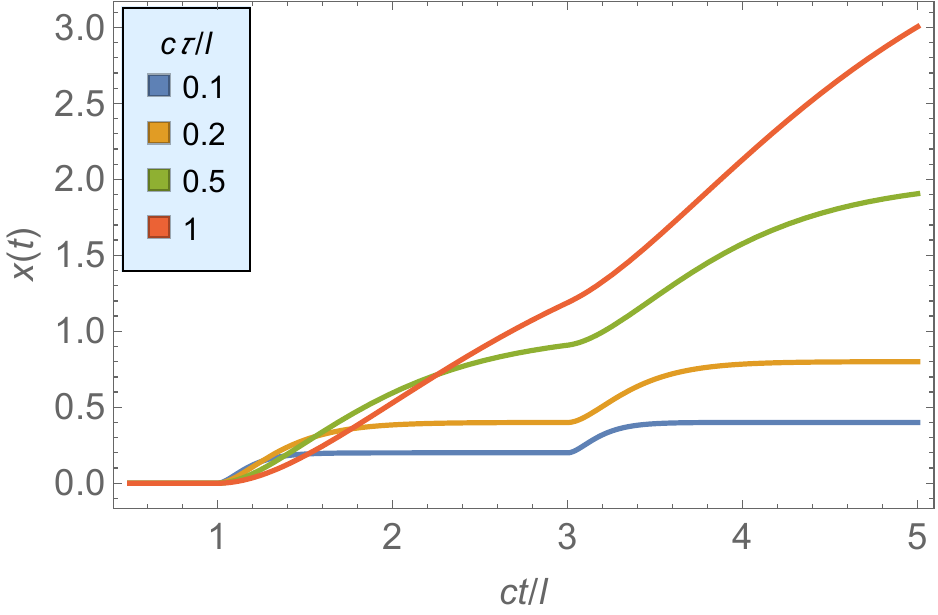}
\caption{
Dependence of the left end displacement $x$ on time $t$.
Calculation using formula \eqref{05.1:4}. The $c\tau/l=0.1$ and $c\tau/l=0.2$ variants approximately correspond to the signals in Fig.~\ref{fig:Astashynski+2014PPT_1_157-Fig2}.
}\label{fig:Example1X-2}
\end{figure}
\begin{figure}
\centering
\includegraphics[width=\linewidth]{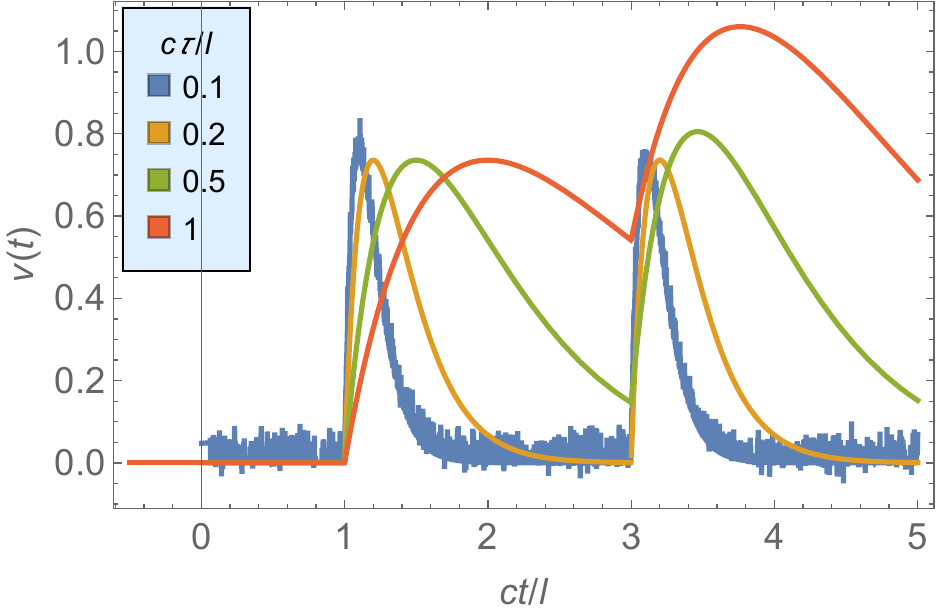}
\caption{
Left-hand end velocity $v$ versus time $t$.
Calculated using formula \eqref{05.1:6}. The spikes in the graph for the $c\tau/l=0.1$ case indicate insufficient precision with the 15-bit arithmetic that Wolfram Mathematica\textsuperscript{\circledR} uses by default.
}\label{fig:Example1V}
\end{figure}

%
%
%

As an example, consider a pressure pulse of the form
\begin{gather}
\label{05.1:1}
f(t)
=
\frac{t}{\tau}
\exp\left[
-\frac{t}{\tau}
\right]
\theta(t)
,
\end{gather}
where the Heaviside theta function is defined as
\begin{gather}
\theta(t)
=
\begin{cases}
0, & \mbox{if } t<0 \\
1, & \mbox{if } t>0 \\
\texttt{undefined}, & t=0.
\end{cases}
\end{gather}
Function \eqref{05.1:1} approximately reproduces the pulse shape measured in Ref.~\cite{Astashynski+2014PPT_1_157}. Its peculiarity is that the force increases smoothly linearly from zero at the initial moment of time and then smoothly decreases approximately exponentially. The Laplace transform results in the simple formula
\begin{gather}
\label{05.1:2}
F(s)
=
\frac{\tau}{(1+s\tau)^{2}}
\end{gather}
for force image. Now, taking into account the transition to dimensionless variables, substituting \eqref{05.1:2} into \eqref{05:11} yields
\begin{multline}
\label{05.1:3}
x(t)
=
\tau
\left(
i \underset{m=1}{\overset{\infty }
{\sum }}
\left(
\frac{\e^{i m \pi\, ( t +1)}}{m \pi\, (m \pi \tau -i)^2}
-
\frac{\e^{-i m \pi \,( t +1)}}{m \pi\, (m \pi \tau +i)^2}
\right)
+
\right.
\\
\left.
+
\frac{\e^{- t / \tau }}{\tau}
\text{csch}\left({1}/\tau \right)
\left( t +\tau +\coth \left({1}/\tau \right)\right)
+ t-2\tau \right)
.
\end{multline}
Wolfram Mathematica\textsuperscript{\circledR}, with some assistance, expresses an infinite sum through special functions:
\begin{multline}
\label{05.1:4}
x( t )
=
\tau
\left\{
\frac{1}{\pi ^2\tau }
\left[
i \pi \e^{i \pi( t +1)} \tau
\Phi\left(
\e^{i \pi( t +1)},1,\frac{\pi \tau -i}{\pi \tau } 
\right) 
+ 
\right. 
\right. 
\\ 
\left. 
\left. 
+ 
\e^{i \pi ( t +1)} 
\Phi\left( 
\e^{i\pi( t +1)},2,\frac{\pi \tau -i}{\pi \tau } 
\right) 
+i \pi \tau \log\left(1-\e^{i \pi ( t +1)}\right) 
\right] 
+ 
\right. 
\\ 
\left. 
+ 
\frac{\e^{-i \pi ( t +1)}}{\pi^2\tau } 
\left[ 
-i\pi\tau 
\Phi\left( 
\e^{-i \pi( t +1)},1,\frac{\pi \tau +i}{\pi\tau } 
\right) 
+ 
\right. 
\right. 
\\ 
\left. 
\left. 
+ 
\Phi\left( 
\e^{-i \pi( t +1)},2,\frac{\pi\tau +i}{\pi\tau } 
\right) 
- 
\right. 
\right. 
\\ 
\left. 
\left. 
- 
i\pi \e^{i\pi( t +1)} \tau 
\log\left( 
\e^{-i \pi ( t +1)}\left(-1+\e^{i \pi( t +1)}\right) 
\right) 
\right] 
+ 
\right. 
\\ 
\left. 
+ 
{1}/\tau 
\left[ 
\e^{-\frac{ t }{\tau }}
\text{csch}
\left({1}/\tau \right)
\left(
t +\tau +\coth \left({1}/\tau \right)
\right)
\right]
+
t -2 \tau
\right\}
,
\end{multline}
where
\begin{gather*}
\Phi(z,s,a)
=
\sum_{k=0}^{\infty }
z^{k}\left(
k+a
\right)^{-s}
\end{gather*}
denotes the Hurwitz-Lerch transcendent. For the velocity of the left end face, we obtain a slightly simpler formula:
\begin{multline}
\label{05.1:6}
v(t)
=
\tau
\left\{ 
\frac{\e^{-i \pi t }}{\pi ^2 \tau ^2} 
\left[ 
\Phi\left(-\e^{-i \pi t },2,1+\frac{i}{\pi\tau }\right) 
+ 
\right. 
\right. 
\\ 
\left. 
\left.
+ 
\e^{2 i\pi t } 
\Phi\left(-\e^{i \pi t },2,1-\frac{i}{\pi\tau }\right) 
\right] 
- 
\right. 
\\ 
\left. 
- 
\frac{1}{\tau ^2} 
\e^{- t / \tau } 
\text{csch}\left({1}/{\tau}\right) 
\left( 
t +\coth \left({1}/\tau \right) 
\right) 
+ 
1 
\right\} 
. 
\end{multline}
The results of calculations using formulas \eqref{05.1:4} and \eqref{05.1:6} are shown in figures \ref{fig:Example1X-2} and \ref{fig:Example1V}. These figures demonstrate that the calculation accuracy of special functions is not always sufficient to obtain a satisfactory result. Looking ahead, we note that these formulas provide only one of many, and not the best, representation of the calculated displacement and velocity of the left end face.

%

\subsection{Solution as a Convolution}
\label{s5.2}

Let's try to represent the found solution as a convolution, for example,
\begin{gather}
\label{05.2:01}
v(t) = \int_{0}^{t} g_{v}(t-t')\,f(t')\dif{t'}
,
\end{gather}
where $g_{v}(t-t')$ is called the transition function, as is customary in radio engineering, or the Green's function, as in the theory of integral equations. Equality \eqref{05.2:01} is called the Volterra equation if the unknown function is the force $f(t')$, not the velocity $v(t)$. It is useful to note that $v(t)=0$ for $t<1$ due to the delay in signal propagation from the right end, where force $f(t)$ begins to act at time $t=0$, to the left end, where its displacement under the action of the incident and reflected waves is measured. Another fact worthy of mention is the agreement of the Volterra equation with the principle of causality. Indeed, according to \eqref{05.2:01}, the velocity $v(t)$ (the response to the force) depends on the force function $f(t')$ (the cause of the response) only at previous times $t'<t$.

By the convolution theorem (see, for example, \cite{Kolokolov+2013}), the image $V(s)$ of the velocity function of the left end of the rod is equal to the product of the images of the Green's function $G_{v}(s)$ and the force $F(s)$, that is,
\begin{gather}
\label{05.2:02}
V(s) = G_{v}(s)\,F(s)
.
\end{gather}
As follows from the previous analysis, namely from formula \eqref{05:09}, the image of the Green's function is
\begin{gather}
\label{05.2:03}
G_{v}(s)
=
\frac{1}{\sinh(s)}
.
\end{gather}
Performing the inverse Laplace transform, we obtain
\begin{gather}
g_{v}(t-t')
=
2 \sum_{j=0}^{\infty} \delta\left(t-t'-1 -2j\right)
.
\end{gather}
Accordingly,
\begin{gather}
\label{05.2:05}
v(t)
=
2
\sum_{j=0}^{\infty}
f(t-1-2j)
.
\end{gather}

Since $f(t-1-2j)=0$ for $t< 1+2j$, the formally infinite series on the right-hand side of formula \eqref{05.2:05} at any finite time actually contains a finite number of terms. To express this fact explicitly, the upper limit of summation in the formula \eqref{05.2:05} should be specified:
\begin{gather}
\label{05.2:05'}
v(t)
=
2
\sum_{j=0}^{\left\lfloor\left(t-1\right)/2\right\rfloor}
f\left(t-1-2j\right)
.
\end{gather}
where $\left\lfloor\left(t-1\right)/2\right\rfloor$ is the largest integer less than $\left(t-1\right)/2$.

By calculating the difference $v\left(t+1\right)-v\left(t-1\right)$, we find that all terms on the right, except the very first, cancel out. This allows us to find a solution to the inverse problem, that is, express the force in terms of velocity:
\begin{gather}
\label{05.2:06}
f(t)
=
\tfrac{1}{2}
\left[
v\left(t+1\right)
- v\left(t-1\right)
\right]
.
\end{gather}
In dimensional notation, this result has the form
\begin{gather}
\label{05.2:06a}
f(t)
=
\tfrac{1}{2l}\,m\,c^{2}
\left[
v\left(t+1\right)
- v\left(t-1\right)
\right]
.
\end{gather}
Converted to pressure, we have the result
\begin{gather}
\label{05.2:07}
p(t)
=
\tfrac{1}{2}\,\rho\,c
\left[
v(t+l/c)-v(t-l/c)
\right]
,
\end{gather}
convenient for comparison with the intuitive formula \eqref{01:01}, which we promised to refute.

When interpreting formula \eqref{05.2:06a}, we must remember that $v(t)=0$ for $t<l/c$. Therefore, in the interval $0<t<2l/c$, the second term is zero, so that
\begin{gather}
\label{05.2:08}
f(t)
=
\tfrac{1}{2l}\,mc\,v(t+l/c)
\qquad
[0<t<2l/c]
.
\end{gather}
In the paper \cite{Astashynski+2014PPT_1_157}, the length of the sensor rod was deliberately increased in order to get rid of the second term, but even then the formula given there turned out to be incorrect due to the absence of the factor $1/2$. As noted above, the origin of this factor in formulas \eqref{05.2:06}, \eqref{05.2:07}, and \eqref{05.2:08} can be explained by the fact that the end-face displacement velocity is the sum of the displacement velocities of the incident and reflected sound waves. The process of propagation of incident and reflected waves is discussed in more detail in Appendix \ref{A1}.

\begin{figure}
\centering
\includegraphics[width=\linewidth]{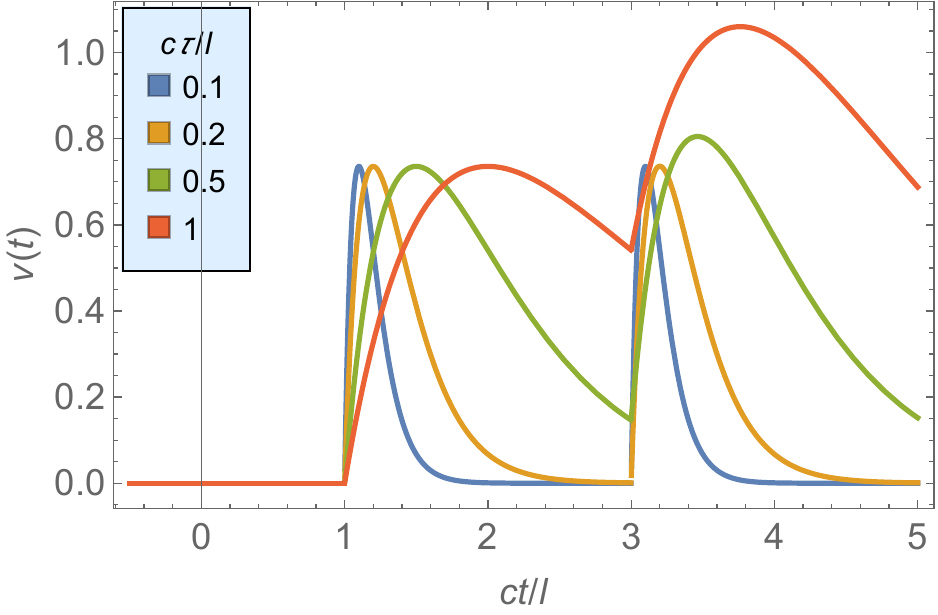}
\caption{
Left end displacement velocity $v(t)$ versus time $t$.
Calculated using formula \eqref{05.2:09}. The spikes visible in Fig.~\ref{fig:Example1V} are absent.
}\label{fig:Example3V}
\end{figure}

\begin{figure}
\centering
\includegraphics[width=\linewidth]{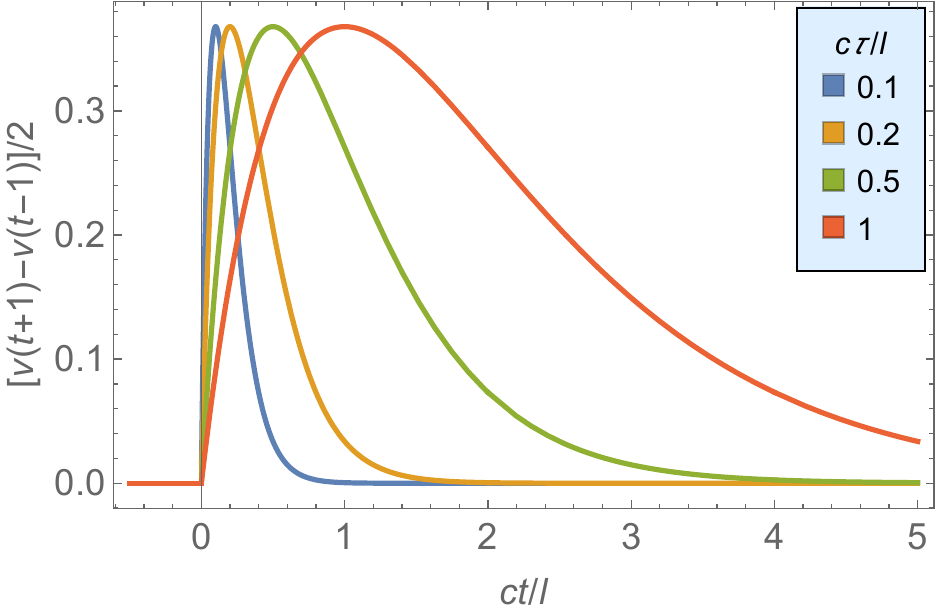}
\caption{
Time $t$ dependence of the force $f(t)$ applied to the right end of the rod, reconstructed from the left end displacement velocity $v(t)$. Calculation using formulas \eqref{05.2:06} and \eqref{05.2:09}.
}\label{fig:Example3V-F}
\end{figure}

Let us also clarify that, for the chosen left-to-right coordinate axis direction, the displacement velocity of the left end of the rod is negative (at least at the first instants of time); accordingly, the force $f(t)$ is also negative. However, when constructing figures \ref{fig:Example1X-2}–\ref{fig:Example3V-F}, it was assumed that $f(t)>0$.

It may seem that formula \eqref{05.2:06a} violates the principle of causality, since it expresses the force (i.e., the cause) at time $t$ in terms of the velocity (the result of the force's action) not only at time $t+l/c$ after the arrival of the force impulse, but also at time $t-l/c$ before it. In fact, in boundary value problems with initial conditions, compliance with the principle of causality is not an absolute requirement for the correctness of the solution. Specifically, in the problem under consideration, due to the reflection of sound waves from the ends of the rod, the velocity $v\left(t-l/c\right)$ at previous times can be expressed in terms of the velocity at subsequent times.

Formula \eqref{05.2:05} somewhat devalues the effort of deriving the formulas in section \ref{s5.1}. Formal summation using this formula for the example in that section gives an alternative representation of the solution found there \eqref{05.1:6}:
\begin{multline}
\label{05.2:09}
v(t)
=
{2 \e^{({1-t})/{\tau }} }
\left\{ 
2 \e^{3/{\tau }-2 \left\lceil{1}/{2}-{t}/{2}\right\rceil/{\tau } }
\right.
\\
\left. 
\times 
\left[ 
\left(2 \left\lceil {1}/{2}-{t}/{2}\right\rceil +t-2\right) 
\sinh \left({1}/{\tau }\right) 
+ 
\cosh \left({1}/{\tau }\right) 
\right]
\right.
\\
\left. 
-
\e^{2/\tau }(t+1)+t-1
\right\}
/
\left[
{\left(\e^{2/\tau }-1\right)^2 \tau }
\right]
,
\end{multline}
\begin{gather}
\label{05.2:10}
x(t) = \int_{1}^{t} v(t')\dif{t'}
,
\end{gather}
if $t>0$, and $v(t)=x(t)=0$ if $t<0$. Calculations using formulas \eqref{05.2:09} and \eqref{05.2:10} take less time and are more accurate, as can be seen from a comparison of Fig. ~\ref{fig:Example1V} and \ref{fig:Example3V}.

\section{Inverse Problem}
\label{s6}

According to the experimental conditions, the measured value is the displacement of the left end of the rod
\begin{gather}
\label{06:01}
x(t)
=
\frac{1}{2\pi i}
\int\limits_{-i\infty +\gamma }^{i\infty +\gamma }
\e^{st}X(s)
\dif{s}
\end{gather}
while the force $f(t)$ must be reconstructed from the measured dependence $x(t)$ or $v(t)=\dot{x}(t)$.


The problem of restoring the force $f(t)$ is the inverse of calculating $x(t)$ and $v(t)$. The inverse problem is often classified as a mathematically ill-posed problem. There are three signs of ill-posedness. A well-posed problem is one for which the following conditions are satisfied \cite{Parker1989}:
\begin{enumerate}
\item
the problem has a solution for any admissible initial conditions (the existence of a solution);

\item
each initial condition corresponds to only one solution (uniqueness of the solution);

\item
the behavior of the solution changes continuously as the initial conditions change (solution stability).

\end{enumerate}
Problems that are not well-posed in the above sense are called ill-posed. Inverse problems are often ill-posed; for example, the inverse heat equation, which derives the initial temperature distribution from the final data, is ill-posed because the solution is very sensitive to changes in the final data. There are examples where the inverse problem has no solution at all. \footnote{
See, for example, \href{https://math.stackexchange.com/questions/380720/is-deconvolution-simply-division-in-frequency-domain}{Is deconvolution simply division in frequency domain?}
Here is a basic explanation of why the inverse operation of convolution is ill-posed.
}
In most physically significant physical problems, the solution to the inverse problem is unstable with respect to measurement errors, that is, the errors in the measured function increase exponentially in the reconstructed function. Such problems can usually be solved using various regularization methods  \cite{TikhonovArsenin1977}. By coincidence, in our case, the inverse problem has already been solved at the end of section \ref{s5.2}. However, we do not yet know how stable the found solution is with respect to measurement errors. Moreover, the primary quantity measured by the modulation of the laser radiation intensity is the displacement $x(t)$ of the left end, whereas the displacement velocity $v(t)$ is obtained from $x(t)$, essentially by difference differentiation, which in itself is a mathematically ill-posed problem. Indeed, numerical differentiation involves composing fractions of the form $(x(t+\Delta t/2)-x(t-\Delta t/2))/\Delta t$, which, as $\Delta t$ tends to zero, have no definite limit due to the inaccuracy of the measurement of $x(t)$.

To clarify these issues, we will try to find a solution to the inverse problem, so to speak, using a universal method, keeping in mind the generalization of the previously found solution in the future to the more complex case of a measuring rod on suspensions.

We will continue to use dimensionless variables and rewrite the convolution \eqref{05.2:02} as
\begin{gather}
\label{06:02}
F(s) = H_{v}(s) \, V(s)
,
\end{gather}
assuming that
\begin{gather}
\label{06:01}
H_{v}(s) = 1/G_{v}(s)
.
\end{gather}
In the example with a free rod
\begin{gather}
\label{06:03}
H_{v}(s) = \sinh(s)
.
\end{gather}
The same result follows directly from equation \eqref{05:09}.
Mellin integral
\begin{gather}
\label{06:04h}
h_{v}(t) = \frac{1}{2\pi i}\int_{-i\infty +\gamma }^{i\infty + \gamma } H_{v}(s) \e^{s t}\dif{s},
\end{gather}
for such a range, the Green's function diverges, meaning the inverse Laplace transform is impossible.

We apply the Mellin integral directly to equation \eqref{06:02}:
\begin{multline*} 
f(t) = \frac{1}{2\pi i}\int_{-i\infty +\gamma }^{i\infty + \gamma } F(s) \e^{s t}\dif{s} 
= 
\\ 
= \frac{1}{2\pi i}\int_{-i\infty +\gamma }^{i\infty + \gamma } H_{v}(s)\,V(s) \e^{s t}\dif{s} 
\\ 
= \frac{1}{2\pi i}\int_{-i\infty +\gamma }^{i\infty + \gamma } H_{v}(s)\,\left( 
\int_{0}^{\infty }v(t')\e^{-st'}\dif{t'}
\right) \e^{s t}\dif{s}
.
\end{multline*}
In the integral over $t'$, we take into account that $v(t')=0$ for $t'<1$, so in the lower limit of integration, zero can be replaced by one without changing the result of integration:
\begin{multline*}
f(t)
=
\frac{1}{2\pi i}\int_{-i\infty +\gamma }^{i\infty + \gamma } H_{v}(s)\,\left(
\int_{1}^{\infty }v(t')\e^{-st'}\dif{t'}
\right) \e^{s t}\dif{s}
\\
=
\frac{1}{2\pi i}\int_{-i\infty +\gamma }^{i\infty + \gamma } H_{v}(s)\,\left(
\int_{0}^{\infty }v(1+t'')\e^{–s-st''}\dif{t''}
\right) \e^{s t}\dif{s}
.
\end{multline*}
Here and below, symbols with a hat denote functions on the shifted time scale, which count time from the moment the sound wave arrives at the left end. Let, in particular,
\begin{gather}
\hat{v}(t'')=v(1+t'')
.
\end{gather}
We will also mark the images of functions of shifted time with a hat.
Then 
\begin{multline*} 
f(t) 
= 
\frac{1}{2\pi i}\int_{-i\infty +\gamma }^{i\infty + \gamma } H_{v}(s)\e^{-s}\,\left( 
\int_{0}^{\infty }\hat{v}(t'')\e^{-st''}\dif{t''} 
\right) \e^{s t}\dif{s} 
\\ 
= 
\frac{1}{2\pi i}\int_{-i\infty +\gamma }^{i\infty + \gamma } H_{v}(s)\e^{-s}\,\hat{V}(s)\, 
\e^{s t}\dif{s} 
, 
\end{multline*}
where 
\begin{gather*} 
\hat{V}(s) 
= 
\int_{0}^{\infty }\hat{v}(t'')\e^{-st''}\dif{t''} 
. 
\end{gather*}
Introducing the notation
\begin{gather}
\label{06:07}
\hat{H}_{v}(s) = \e^{-s}H_{v}(s) = \e^{-s}/G_{v}(s)
,
\end{gather}
we obtain
\begin{gather}
f(t)
=
\frac{1}{2\pi i}\int_{-i\infty +\gamma }^{i\infty + \gamma }
\hat{H}_{v}(s)\,\hat{V}(s)\,\e^{s t}\dif{s}
,
\\
\intertext{that is}
F(s) = \hat{H}_{v}(s)\,\hat{V}(s) 
. 
\end{gather}
In the case of a free rod
\begin{gather}
\hat{H}_{v}(s)
=
\frac{1}{2}\left(
1-\e^{-s}
\right)
\end{gather}
and
\begin{gather}
\label{06:11}
\hat{h}_{v}(t)
=
\frac{1}{2\pi i}\int_{-i\infty +\gamma }^{i\infty + \gamma }
\hat{H}_{v}(s) \e^{s t}\dif{s}
=
\frac{1}{2}\left[
\delta (t) - \delta (t-2)
\right]
,
\end{gather}
By the convolution theorem we have 
\begin{multline} 
\label{06:12} 
f(t) 
= 
\int_{0}^{t} 
\hat{h}_{v}(t-t')\,\hat{v}(t')\dif{t'} 
= 
\\ 
=\int_{0}^{t} 
\frac{1}{2}\left[ 
\delta (t-t') - \delta (t-t'-2) 
\right] 
v(1+t')\dif{t'} 
\\ 
=\frac{1}{2}\left[ 
v(1+t) - v(t-1) 
\right] 
,
\end{multline}
which reproduces the already known result \eqref{05.2:06}. Extrapolating to the case of a rod with suspensions, which will be considered in a separate article, we arrive at the rule
\begin{gather}
\label{06:14}
f(t)
=
\int_{0}^{t} \hat{h}_{v}(t-t')\,v(1+t')\dif{t'}
,
\end{gather}
where
\begin{gather}
\label{06:15}
\hat{h}_{v}(t-t')
=
\frac{1}{2\pi i}\int_{-i\infty +\gamma }^{i\infty + \gamma }
\e^{s (t-t')}\frac{\e^{-s}}{G_{v}(s)}\, \dif{s}
.
\end{gather}

\begin{figure*}
  \centering
  \includegraphics[width=0.475\linewidth]{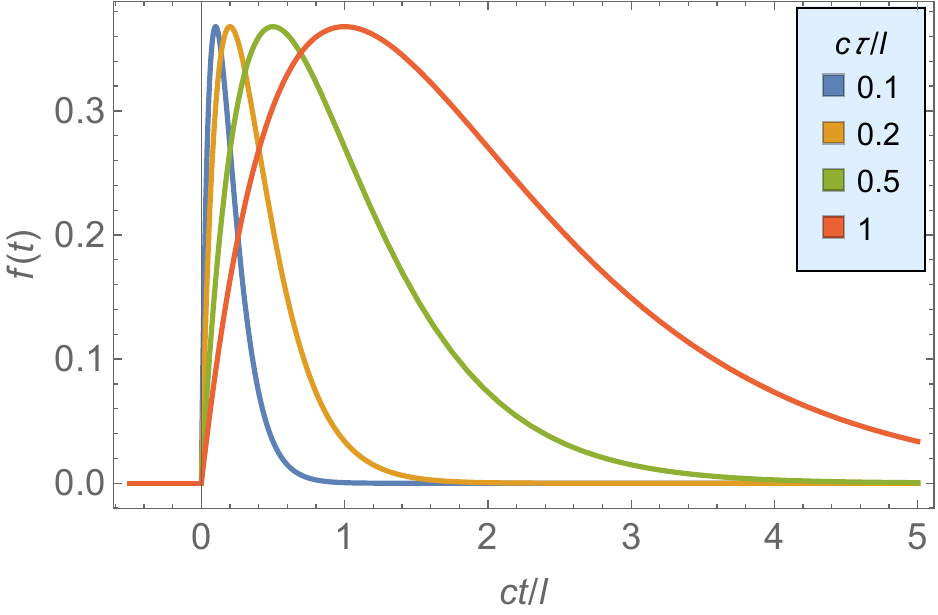}
  \hfil
  \includegraphics[width=0.475\linewidth]{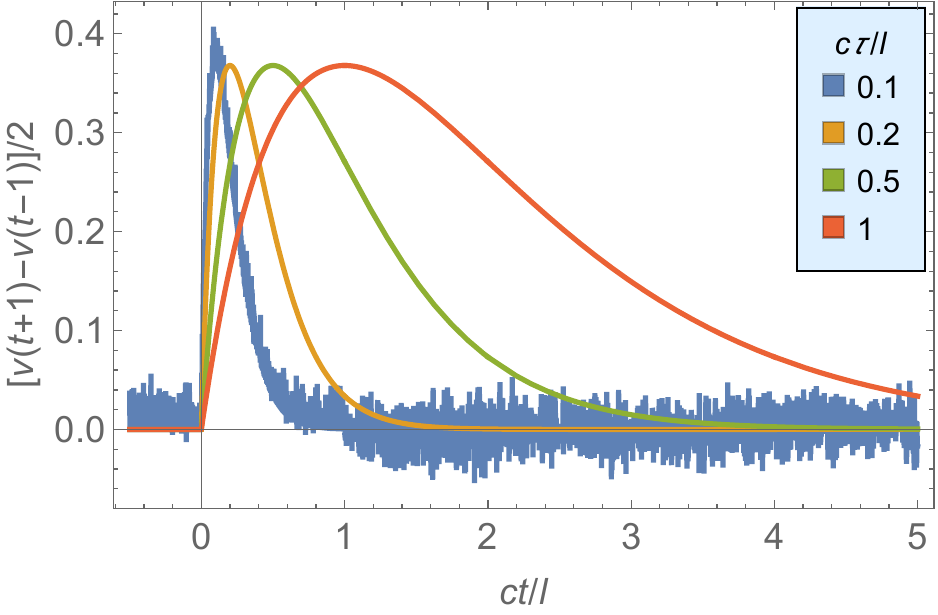}
\caption{
Example 1.
Force \eqref{05.1:1} (left) and its restoration using formulas \eqref{05.2:06} and \eqref{05.1:6} (right).
}  \label{fig:Example1F}
\end{figure*}

The explicit solution \eqref{05.2:06} to the inverse problem allows us to conclude that it is robust to measurement errors. As stated in the Introduction, in section \ref{s1}, the measured quantity is, in fact, the displacement of the left end $x(t)$. The velocity $v(t)$ is found from the function $x(t)$, essentially by its numerical differentiation. As stated above, numerical differentiation should be considered a mathematically incorrect operation.

%
%
%
%
%
%
%

\section{Conclusion}
\label{s9}

The chain of events that led to the publication of this article began with a presentation by the developers of a method for measuring the dynamic pressure of a plasma flow \cite{AstashinskiPenyazkov2024Zvenigorod51_40} at the annual plasma physics conference in Zvenigorod in March 2024. The presentation attracted the interest of one of the authors of this article (EK), who discussed it at the weekly planning meeting of his laboratory at the Budker Institute of Nuclear Physics. A few weeks later, another author of this article (IK) from the same laboratory quite by chance came across a preprint by A. Milstein, a member of the Theoretical Department of the Budker INP, \cite{Milstein2024Arxiv_2405_00050}, devoted to a critique of a student's solution to a problem about the oscillations of a bob on a finite-mass spring. Seeing an analogy with the problem of sound wave propagation in a measuring rod, IK applied Milstein's solution to this problem. The third author of this article (VP) proposed his own solution, described in Appendix \ref{A1}.

The main result of this work can be considered formula \eqref{05.2:07}.

The solution to the more complex problem of measuring plasma flow pressure using a measuring rod attached to suspensions will be described in our next article.

\begin{acknowledgments}
This work is part of a state assignment from the Russian Federation to the  Budker Institute of Nuclear Physics.

The authors express their gratitude to Alexander Milstein for numerous discussions and, in particular, for his clarification of the causality principle in boundary value problems with initial conditions.
\end{acknowledgments}

\section*{ORCID ID}

\noindent
Evgeny KOLESNIKOV \href{https://orcid.org/0000-0001-6535-3040}{https://orcid.org/0000-0001-6535-3040}

\noindent
Igor KOTELNIKOV \href{https://orcid.org/0000-0002-5509-3174}{https://orcid.org/0000-0002-5509-3174}

\noindent
Vadim PRIKHODKO \href{https://orcid.org/0000-0003-0199-3035}{https://orcid.org/0000-0003-0199-3035}

\appendix

\section{Partial Wave Method}
\label{A1}

The general solution of the wave equation \eqref{1:0} can be sought in the form of two waves traveling in opposite directions, with the profile of each wave preserved due to neglecting dispersion in the equation \eqref{1:0}:
\begin{gather}
\label{A1:1}
x(z,t) = \xi_{+}(t-z)+\xi_{-}(t+z)
,
\\
\label{A1:2}
v(z,t) = \xi_{+}'(t-z)+\xi_{-}'(t+z)
.
\end{gather}
Here the prime denotes the derivative with respect to the argument of the functions $\xi_{+}$ and $\xi_{-}$. We will call the wave $\xi_{-}$ incident on the left (mirror) end in Fig.~\ref{fig:Kolesnikov}, and the wave $\xi_{+}$ reflected from this end.

The boundary conditions \eqref{1:3} and \eqref{1:4} generate two equations for the functions $\xi_{+}$ and $\xi_{-}$:
\begin{gather}
\label{A1:4}
-\xi_{+}'(t-1) + \xi_{-}'(t+1)
=
f(t)
,
\\
\label{A1:5}
-\xi_{+}'(t) + \xi_{-}'(t)
=
0.
\end{gather}
The last of these, \eqref{A1:5}, eliminates the need to distinguish between the derivatives of both waves: $\xi_{+}'(t) = \xi_{-}'(t)$. Accordingly, the functions $\xi_{+}(t)$ and $\xi_{-}(t)$, if they differ at all for the same values of their arguments, do so only by a constant, but this constant must be zero to satisfy the initial conditions.

Equation \eqref{A1:2} relates the derivatives $\xi_{+}(t)$ and $\xi_{-}(t)$ to the measured quantity.
\begin{gather}
\label{A1:06}
v(t) = v(0,t) = 2 \xi_{+}'(t) = 2 \xi_{-}'(t).
\end{gather}
Determining $\xi_{+}(t-1)$ and $\xi_{-}(t+1)$ from this and substituting them into \eqref{A1:4}, we obtain equation \eqref{05.2:06}. If we rewrite it as a recursion
\begin{gather}
v(t) = v(t-2) + 2 f(t-1)
,
\end{gather}
after $n$ iterations, we obtain the relation
\begin{gather}
\label{A1:08}
v(t) = 2 \sum_{j=0}^{n-1} f(t-1-2j) + v(t-2n) ,
\end{gather}
in which the last term becomes zero if $t-2n<1$, since $v(t)=0$ for $t < 1$. Thus, we again obtain the formula \eqref{05.2:05'}.


The end face displacement $x(t)$ can be found by directly integrating equation \eqref{05.2:05'}. Introducing the notation
\begin{gather}
\label{A1:09}
\phi(t) = \int_{-\infty}^{t} f(t') \dif{t}',
\end{gather}
where $f(t')=0$ for $t'<0$, the result can be written as
\begin{gather}
\label{A1:11}
\xi(t) = \sum_{j=0}^{\infty} \phi(t-1-2j)
. \end{gather}
If desired, the upper limit of the summation can be replaced by $\left\lfloor\left(t-1\right)/2\right\rfloor$, as in \eqref{05.2:05'}, to emphasize that at any given time the sum contains a finite number of terms.

Finally, by combining \eqref{A1:1}, \eqref{A1:06}, and \eqref{A1:11}, it is easy to find the displacement of an arbitrary element of the rod at any given time:
\begin{multline}
\label{A1:12}
x(z,t)
=
\sum_{j=0}^{\infty}
\left[
\phi\left(t-{(2j+1)-z}\right)
\right.
\\
\left.
+
\phi\left(t-{(2j+1)+z}\right)
\right]
.
\end{multline}
This series for any finite instant of time contains a finite number of terms and also satisfies the wave equation \eqref{1:0} and the boundary conditions \eqref{1:1}--\eqref{1:4}.


\begin{thebibliography}{10}
\def\selectlanguageifdefined#1{
\expandafter\ifx\csname date#1\endcsname\relax
\else\selectlanguage{#1}\fi}
\providecommand*{\href}[2]{{\small #2}}
\providecommand*{\url}[1]{{\small #1}}
\providecommand*{\BibUrl}[1]{\url{#1}}
\providecommand{\BibAnnote}[1]{}
\providecommand*{\BibEmph}[1]{#1}
\ProvideTextCommandDefault{\cyrdash}{\iflanguage{russian}{\hbox
  to.8em{--\hss--}}{\textemdash}}
\providecommand*{\BibDash}{\ifdim\lastskip>0pt\unskip\nobreak\hskip.2em plus
  0.1em\fi
\cyrdash\hskip.2em plus 0.1em\ignorespaces}
\renewcommand{\newblock}{\ignorespaces}

\bibitem{Kostukevich2002PublAstronObs_74_149}
\selectlanguageifdefined{russian}
Костюкевич~Е.~А. Возможности оптических
  датчиков импульсного давления в
  плазмодинамическом эксперименте~//
  \BibEmph{Publ. Astron. Obs.} \BibDash
\newblock 2002. \BibDash
\newblock {№}~74. \BibDash
\newblock {С.}~149.

\bibitem{Astashynski+2014PPT_1_157}
\selectlanguageifdefined{english}
Studies on Dynamic Pressure of Compression Plasma Flow~/ Astashynski~V.~M.,
  Sari~A.~H., Ananin~S.~I., Kostyukevich~E.~A., Shoronov~P.~N., and
  Kuzmitski~A.~M.~// \BibEmph{Plasma Physics and Technology}. \BibDash
\newblock 2014. \BibDash
\newblock Vol.~1, no.~3. \BibDash
\newblock P.~157--159.

\bibitem{IvanovPrikhodko2017PhysUsp_60_509}
\selectlanguageifdefined{english}
Ivanov~A.~A., Prikhodko~V.~V. Gas dynamic trap: experimental results and future
  prospects~//
  \href{https://doi.org/10.3367/UFNe.2016.09.037967}{\BibEmph{Phys. Usp.}}
  \BibDash
\newblock 2017. \BibDash
\newblock Vol.~60, no.~5. \BibDash
\newblock P.~509--533. \BibDash
\newblock Access mode: \BibUrl{https://ufn.ru/en/articles/2017/5/d/}.

\bibitem{Morozov2008}
\selectlanguageifdefined{russian}
Морозов~А.~И. Введение в плазмодинамику.
  \BibDash
\newblock 2е изд. {изд.} \BibDash
\newblock М.~: Физматлит, 2008. \BibDash
\newblock
  ISBN:~\href{http://isbndb.com/search-all.html?kw=978-5-9221-0931-4}{978-5-9221-0931-4}.

\bibitem{ManzhosovMartynova2021VestnikUlTU_3_86}
\selectlanguageifdefined{russian}
Манжосов~В., Мартынова~Н.~Б. Движение
  однородного стержня при действии
  постоянного давления на торце~//
  \BibEmph{Вестник Ульяновского
  государственного технического
  университета}. \BibDash
\newblock 2001. \BibDash
\newblock {№} 3 (15). \BibDash
\newblock {С.}~86--91.

\bibitem{LLVI}
\selectlanguageifdefined{russian}
Ландау~Л.~Д., Лифшиц~Е.~М. Теоретическая
  физика: учеб. пособие в 10~т. \BibDash
\newblock 3 {изд.} \BibDash
\newblock М.~: Наука, 1986. \BibDash
\newblock Т.~VI. Гидродинамика. \BibDash
\newblock 720~{с.}

\bibitem{Milstein2024Arxiv_2405_00050}
\selectlanguageifdefined{english}
Milstein~A.~I. Non-trivial solution to a simple problem. \BibDash
\newblock 2024. \BibDash
\newblock 2405.00050.

\bibitem{LLVII}
\selectlanguageifdefined{russian}
Ландау~Л.~Д., Лифшиц~Е.~М. Теоретическая
  физика: учеб. пособие в 10~т. \BibDash
\newblock 4 {изд.} \BibDash
\newblock М.~: Наука, 1987. \BibDash
\newblock Т.~VII. Теория уаругости. \BibDash
\newblock 247~{с.}

\bibitem{Kolokolov+2013}
\selectlanguageifdefined{russian}
Задачи по математическим методам физики~/
  Колоколов~И.В.,
  Кузнецов~Е.А., Мильштейн А.И.,
  Подивилов~Е.В.,
  Черных~А.И.,
  Шапиро~Д.~А. и
  Шапиро~Е.Г. \BibDash
\newblock Москва~: URSS, 2013. \BibDash
\newblock 286~{с.} \BibDash
\newblock
  ISBN:~\href{http://isbndb.com/search-all.html?kw=978-5-397-03680-1}{978-5-397-03680-1}.

\bibitem{Parker1989}
\selectlanguageifdefined{english}
McGraw-Hill Dictionary of Scientific and Technical Terms~/ ed.\ by\
  Parker~Sybil~B. \BibDash
\newblock 4th ed. \BibDash
\newblock New York~: McGraw-Hill, 1989. \BibDash
\newblock
  ISBN:~\href{http://isbndb.com/search-all.html?kw=0-07-045270-9}{0-07-045270-9}.

\bibitem{TikhonovArsenin1977}
\selectlanguageifdefined{english}
Tikhonov~A.~N., Arsenin~V.~Y. Solutions of ill-Posed Problems. \BibDash
\newblock New York~: Winston, 1977. \BibDash
\newblock
  ISBN:~\href{http://isbndb.com/search-all.html?kw=0-470-99124-0}{0-470-99124-0}.

\bibitem{AstashinskiPenyazkov2024Zvenigorod51_40}
\selectlanguageifdefined{russian}
Асташинский~В.~М., Пенязьков~О.~Г.
  \href{https://doi.org/10.34854/ICPAF.51.2024.1.1.004}{Квазистационарные
  плазмодинамические системы в науке и
  технологиях}~// 51-я конференция по физике
  плазмы и управляемому термоядерному
  синтезу, 18—22 марта 2024 года~/ РОСАТОМ, РАН.
  \BibDash
\newblock Москва~: ООО «Издательство МБА». \BibDash
\newblock 2024. \BibDash
\newblock {С.}~40. \BibDash
\newblock {Режим доступа}:
  \BibUrl{http://www.fpl.gpi.ru/Zvenigorod/LI/Sbornik-2024.pdf}.

\end{thebibliography}

%
%
%

\end{document}